\documentclass[twocolumn,tightenlines,showpacs,aps]{revtex4}
\usepackage{graphicx}

\begin{document}

\title{Temperature-dependent Raman scattering of natural and isotopically
substituted PbS}
\author{P. G. Etchegoin$^{a}$}
\author{M. Cardona$^{b}$}
\author{R. Lauck$^{b}$}
\author{R. J. H. Clark$^{c}$}
\author{J. Serrano$^{d}$}
\author{A. H. Romero$^{e}$}
\affiliation{$^{a}$The MacDiarmid Institute for Advanced Materials and Nanotechnology\\
School of Chemical and Physical Sciences\\
Victoria University of Wellington\\
PO Box 600 Wellington, New Zealand}
\affiliation{$^{b}$Max-Planck Institut f{\"u}r Festk{\"o}rperforschung\\
Heisenbergstra\ss{}e 1, D-70569 Stuttgart, Germany}
\affiliation{$^{c}$Christopher Ingold Laboratories, Department of Chemistry, University
College London\\
20 Gordon Street, London WC1H 0AJ, United Kingdom}
\affiliation{$^{d}$European Synchrotron Radiation Facility
Boite Postale 220, 38043 Grenoble, France}
\affiliation{$^{e}$CINVESTAV-Quer\'etaro, Libramiento Norponiente 2000, 76230, Fracc.
Real de Juriquilla, Quer\'{e}taro, M\'exico}
\date{\today}

\begin{abstract}
Lead sulfide is an important semiconductor that has found technological
applications for over a century. Raman spectroscopy, a standard tool for the
investigation and characterization of semiconductors, has limited
application to this material because of the forbidden nature of its first
order scattering and its opacity to visible lasers. Nevertheless, useful
vibrational spectra from two-phonon processes are obtained with red lasers,
probably because of a resonance in the concomitant electronic transitions.
Here we report temperature dependent spectra, covering the 10-300\thinspace
K range, for two samples with different sulfur isotopic compositions. The
results are analyzed by comparison with {\it ab initio} calculations of
the lattice dynamics of PbS and the corresponding densities of one and
two-phonon states. Emphasis is placed on the analysis of the two phonon band
centered at $\sim 430\,\mathrm{cm}^{-1}$.
\end{abstract}

\maketitle

%\pacs{}

\section{Introduction}

Lead sulfide (PbS), the mineral galena, is an important material in lead
smelting\cite{1}. It can be found in nature and also grown as an n- or p-type
semiconductor with carrier concentrations as low as $10^{17}\,\mathrm{cm}%
^{-3}$. Its small electronic gap (0.4\thinspace eV), with an anomalous
temperature dependence\cite{2}, implies small effective masses which make
PbS useful as a material for nanostructures\cite{3}. On the negative side,
it is a visually offensive black product which appears during the
degradation of lead white pigment in artwork\cite{4}. It has been used as a
nanoparticle source in very ancient metallic lustre pottery, possibly the
oldest known form of controlled nanotechnology\cite{5}. It has also found an age
old application in the adulteration of an eye-lining cosmetic called
``kohl'' (Sb$_2$S$_3$)\cite{6}.

Raman spectroscopy provides a versatile, non-destructive technique for the
investigation and characterization of semiconducting materials\cite{book}.
The NaCl structure of PbS, however, makes the material first order Raman
forbidden. Nevertheless, it has been recently discovered that characteristic
Raman spectra are obtained when lasers in the red are used as Raman
spectroscopic sources. These spectra are dominated by a band (actually a
doublet) centered at around $\sim 430\,\mathrm{cm}^{-1}$ and a somewhat
narrower peak at $\sim 210\,\mathrm{cm}^{-1}$. A few other characteristic
weak structures are also observed. These spectra have been recently used for
the characterization of lustre pottery and possibly adulterated ``kohl''\cite%
{5,6}.

In a recent publication\cite{Sherwin}, our group investigated the dependence
of the two aforementioned Raman bands on the isotopic mass of sulfur and on
temperature, the latter being limited to the 100-300\thinspace K region. The
lack of data at lower temperatures hampers the investigation of the
mechanisms responsible for the observed anharmonic temperature shifts.
Therefore, in the present work we undertook the investigation of these Raman
spectra down to a temperature of $\sim 10\,\mathrm{K}$. As in the case of
the anomalous temperature dependence of the electronic gap (but most likely
unrelated to it) the frequency of the $\sim 430\,\mathrm{cm}^{-1}$ band
anomalously increases with increasing temperatures, a fact that suggests
interaction of each of the two phonons responsible for this $\sim 430\,%
\mathrm{cm}^{-1}$ band with a set of two phonons at lower frequencies ($\sim
65\,\mathrm{cm}^{-1}$). The dependence of the $430\,\mathrm{cm}^{-1}$ band
on the isotopic mass of sulfur, however, is normal and suggests that the
corresponding eigenvectors are sulfur dominated (for a general discussion of
the dependence of anharmonic self-energies on isotopic masses see Ref. 9).
In order to complete the assignment of the observed Raman structures, we
have calculated the phonon dispersion relations of PbS based on an {\it ab initio} LDA electronic band structure\cite{addedReference1}. Using these dispersion relations, which are rather close to the
few points measured with inelastic neutron scattering (INS), we have
calculated the density of two phonon states for both sums and differences
of two phonons, which is relevant to the interpretation of the two-phonon
Raman scattering. A thorough discussion of the observed spectra based on the
calculated two phonon density of states is given.

\section{Experimental}

The samples were similar (in some cases identical) to those used in Ref. \onlinecite{Sherwin}. The synthetic samples were grown as discussed in Refs. \onlinecite{Sherwin,7}. The natural PbS samples whose spectra we report were mineral galena
(cleaved surfaces) from Creede, Colorado. They were p-type with hole
concentrations $\allowbreak N_{h}\sim 2 \times 10^{17}\,{\rm cm}^{-3}$. The
isotopically modified samples were also p-type, with $N_{h}\sim 10^{18}\,{\rm cm}^{-3}.$ They were mounted in a closed-cycle He-cryostat
(CTI-Cryogenics, Mansfield, MA USA) with temperature control in the
range $10-300\,\mathrm{K}$ and optical quartz windows. Raman measurements
were performed using the $647\,\mathrm{nm}$ line of a Kr$^{+}$-ion laser
which was focused to a $\sim 50\,\mu \mathrm{m}$ diameter spot with $50\,%
\mathrm{mW}$ incident power measured at the entrance window of the cryostat.
The signal was collected using a photographic zoom lens (Canon, $\times $10
magnification) of high numerical aperture (0.45) onto the entrance slit of a
high-dispersion double-additive U1000 Jobin-Yvon spectrometer coupled to a
liquid N$_{2}$-cooled CCD detector (512 pixels) without any polarization
analysis. Integration times in the range $5-15\,\mathrm{min}$ (repeated 3 times, for the software to identify and discard spurious cosmic rays in the spectra) were
used for different measurements in each window of the spectral range. The
high dispersion of the double U1000 spectrometer in the red region ($\sim
647\,\mathrm{nm)}$ implies that several CCD-windows (eight) are needed to
cover the full spectral range from $150$ to $550\,\mathrm{cm}^{-1}$. Data
acquisition of the intrinsically weak second-order spectra is therefore slow
and ultimately limited by the long-term stability of the optical system plus cryostat. The chosen integration times were a compromise between
signal-to-noise ratio and a total integration time that would allow a full
temperature scan within a period of several hours.

\begin{figure}[h]
\centering{\ \includegraphics[width=7cm, height = 6.75cm]{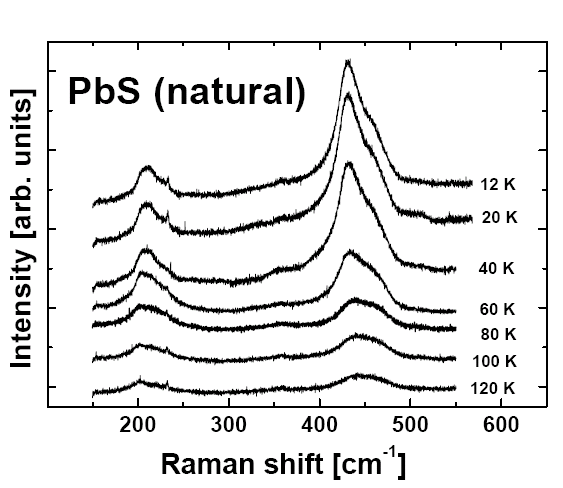} }
\caption{Raw temperature dependence of the second-order Raman scattering of
natural PbS at different temperatures in the range $12-120\, \mathrm{K}$.
The data are for $647\, \mathrm{nm}$ laser excitation.}
\label{fig1}
\end{figure}

Figure \ref{fig1} shows our raw second-order Raman data for natural PbS at
different temperatures. These spectra are dominated by two main features in
the $\sim 210$ and $\sim 430\,\mathrm{cm}^{-1}$ regions. There is evidence
in the experimental spectra for substructure in these peaks, to wit: $(i)$
The singularity at $\sim 430\,\mathrm{cm}^{-1}$ is clearly a doublet and can
be represented as such throughout the entire temperature range by means of
two gaussians. $(ii)$ The singularity at $\sim 210\,\mathrm{cm}^{-1}$ only
reveals itself as a broad doublet at temperatures below $\sim 30\,\mathrm{K}$%
, as we shall show in what follows. Above $\sim 30\,\mathrm{K}$ it is very
difficult to isolate its two individual components, most likely because of
increasing anharmonic broadening\cite{review}.

As already mentioned, the $647\,{\rm nm}$ Kr$^{+}$-laser line was one of the
few lines with which data could be obtained in both natural and isotopically
substituted PbS. Going farther into the visible (in the green, for example,
at 514\,nm, see Fig. 2 of Ref. \onlinecite{7}) resulted in undesirable
(extrinsic) backgrounds that almost completely obliterated the (already very
weak) second-order spectra (see also Fig. 8 of Ref. \onlinecite{Cao}). On the other hand, going into
the near-IR region (e.g. the $\allowbreak 676$\thinspace\ nm line of a Kr$%
^{+}$-ion laser) results in very weak intrinsic second-order spectra. Laser
excitation around $647\,\mathrm{nm}$ seemed to be a good compromise between
low background and reasonable signal-to-noise level (possibly aided by a
weak resonance effect in the red \cite{7}). Another important experimental
detail is the power level of the laser. The surface of PbS
tends to be rather sensitive to photodecomposition (mainly, but not
exclusively, in the presence of oxygen). The power densities used in these
measurements are well below any measurable photo-degradation of the surface
throughout the total integration time.

The analysis presented here is based on the observation of two main
structures in the Raman spectra: $(i)$ one structure at high wavenumbers,
around $\sim 430\,\mathrm{cm}^{-1},$ which is a doublet and has already been
related to the generation of two optical (actually $LO$) phonons \cite{Sherwin,7} and, $(ii)$ a second structure at $\sim 210\,\mathrm{cm}^{-1},$
which is possibly also a doublet but cannot be well resolved in most of the
measured spectra and must be related to either two acoustic phonons ($2LA$) or two
optic ones ($2TO$) as will be discussed later. Our experimental approach is to
characterize the behavior of these spectral features as a function of
temperature and to ascertain how they differ between natural and
isotopically substituted samples. Beside these main features, some weak and
broad structures have been surmised at $\sim 160$, 340, and 500\,cm$^{-1}$. These structures probably correspond to those in Fig. 4 of Ref. \onlinecite{Cao}
at 174, and 320\,cm$^{-1}$. In this figure, a rather sharp peak appears at
135\,cm$^{-1}$ which is below the frequency range of our spectra and we
cannot definitely assign to PbS. The peaks observed in Fig. 4 of Ref. \onlinecite{Cao} at 600 and 961\,cm$^{-1}$ do not seem to correspond to the second-order spectrum of PbS. They are probably due to the $\nu_1$ and
$\nu_4$ vibrations of SO$_4^{-2}$ tetrahedra in some form of lead sulfate\cite{Herzberg,Robin}.

Resolving the different peaks which contribute to the main two singularities
under discussion is the first step to understand their temperature dependent
behavior and, accordingly, their dependence on isotopic mass. The basic
limitation here is the lack of a reliable anaytical model for the
singularities in the phonon densities of states (see next section)
responsible for the second-order scattering observed. Gaussian
lineshapes represent the individual singularities rather well, and will be
used throughout our analysis. The fact that Gaussian lineshapes work well
may also indicate that the shape of the observed structures is heavily
influenced by anharmonicity, rather than by the specific type of DOS
critical points involved. An additional problem is the reliability of the
fits at high temperatures, where even the high energy doublet around $\sim
430\,\mathrm{cm}^{-1}$ merges into a visually unresolved single peak, and
the signal-to-noise ratio worsens considerably. Fits can still be obtained
by restricting the range in which the different parameters are allowed to
vary. Resolution of the two components, however, is not a problem at
temperatures below $\sim 60\,\mathrm{K}$: at these temperatures the $\sim 430\,{\rm cm}^{-1}$ peaks are resolved even to the naked eye.

\begin{figure}[h]
\centering{\ \includegraphics[width=5cm, height = 10cm]{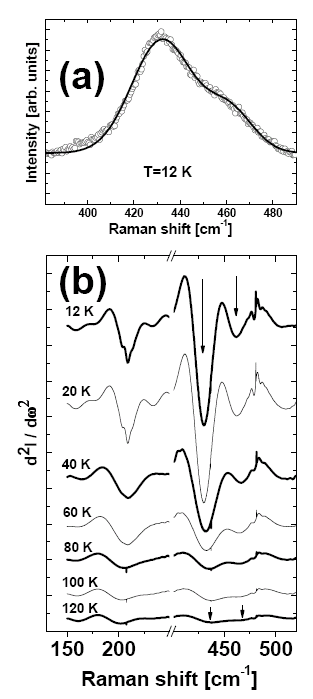} }
\caption{Analysis of the second-order Raman scattering features of natural
PbS. Two approaches have been followed; both giving similar results. In (a)
we show the direct fit of the high energy peak with two Gaussians at $T=12\,%
\mathrm{K}$. The fit has a few imperfections in the lineshape but is an
acceptable representation of the singularity in the density of states. The 
\textit{numerical second derivative} approach to the estimation of critical
point energies is shown in (b) for both singularities (at high and low
wavenumbers) in Fig. \ref{fig1} (note the break in the horizontal axis). The
low wavenumbers ``acoustic'' feature splits into a clear doublet at low
temperatures. The energies of the two peaks in the high energy feature
(obtained from the local minimum in $d^{2}I/d\protect\omega ^{2}$) can be
easily followed throughout the entire temperature range. See the text for
further details. The arrows show the wavenumbers of the bands contributing to the feature at $\sim 430\,{\rm cm}^{-1}$ at both $T=12$ and 120\,K, respectively.}
\label{fig2}
\end{figure}

We tried two ways of isolating the different contributions to the singularities as depicted in Fig. \ref{fig2}.
One is a direct fit with two Gaussians, an example of which is shown in Fig. %
\ref{fig2}(a) for the high-wavenumbers feature. An alternative way that
works satisfactorily throughout the entire temperature range is also to
estimate the energy of the peaks by numerically calculating the second
derivatives of the observed spectra (a common practice when dealing with
electronic dielectric functions and the related energy gaps); such
derivative spectra are shown in Fig. \ref{fig2}(b). In this latter case the
analysis becomes independent of the specific lineshape model used to fit the
data: the critical energies can be assigned to the wavenumber of the
deepest spectral features in the derivative spectra (see arrows in Fig. \ref{fig2}(b)).

\begin{figure}[h]
\centering{\ \includegraphics[width=7cm, height = 6.75cm]{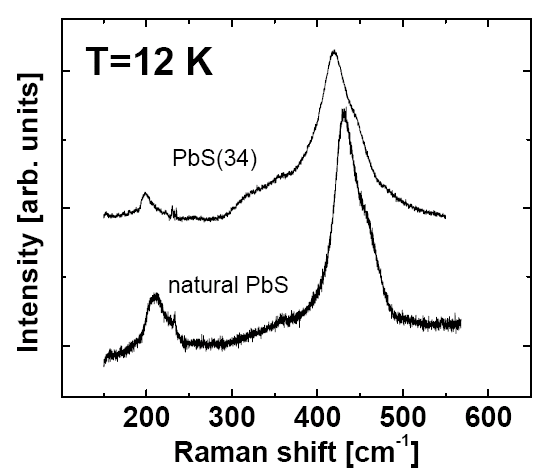} }
\caption{Comparison of the raw Raman second-order spectra between natural and
isotopically substituted Pb$^{34}$S at $12\,\mathrm{K}$. The shift to lower
energies of the singularity at $\sim 430\,\mathrm{cm}^{-1}$ can be easily
seen with the naked eye. In terms of percentages of change in the energies,
the singularity at low wavenumbers ($\sim 210\,\mathrm{cm}^{-1}$) is less
susceptible to a change into a heavier S mass in the isotopically
susbtituted sample.}
\label{fig3}
\end{figure}

The best comparison between natural PbS and isotopically substituted Pb$^{34}$S is obtained from the direct observation of the spectra at the
lowest achievable temperature in our experimental setup (i.e. $12\,\mathrm{K}%
)$, as shown in Fig. \ref{fig3}. The maximum of the low-wavenumbers feature
shifts from $\sim 205\,\mathrm{cm}^{-1}$ in natural PbS to $\sim 201\,%
\mathrm{cm}^{-1}$ in Pb$^{34}$S at $12\,{\rm K}$ ($\sim 2\%$ shift). However, the
corresponding lineshapes are not exactly the same and therefore the shift so
obtained is not expected to be very precise. The maximum of the
high-wavenumbers feature, on the other hand, moves by $\sim 15\,\mathrm{cm}%
^{-1}$, from $432$ down to $417\,\mathrm{cm}^{-1}(3.5\%)$ between the two
samples. Despite differences in the lineshapes between one case and the
other, the feature at high wavenumbers seems to be more susceptible to a
mass change in S than that at low wavenumbers.

The temperature dependence of the low-wavenumbers feature is difficult to
follow for a variety of reasons which include the fact that a doublet cannot
be discerned above $\sim 40\,\mathrm{K}$. In addition, this feature does not
seem to be very sensitive to the sulfur mass substitution. For a discussion of
the dependence of the observed shift with temperature we
therefore concentrate on the high-wavenumber feature at $\sim 430\,\mathrm{cm}%
^{-1}$ . This feature shifts up in frequency with increasing temperatures (an
unusual sign of the shift\cite{review}) Note that the 210 cm$^{-1}$ peak,
and that at 150 cm$^{-1}$, both show the usual down-shift with increasing
temperatures\cite{Sherwin}.

\begin{figure}[h]
\centering{\ \includegraphics[width=7cm, height = 10cm]{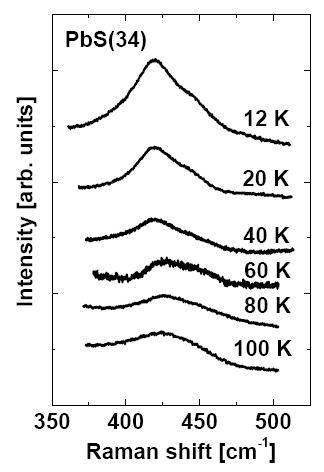} }
\caption{Temperature dependence of the high-wavenumbers feature in the
second-order scattering of Pb$^{34}$S. Both types of analysis of the
spectra, i.e. fits with two Gaussians or numerical second-order derivatives
cast similar results within $\sim 1\, \mathrm{cm}^{-1}$ for the frequencies
of the singularities. The temperature dependence of the low-wavenumbers peak
in this doublet is displayed in Fig. \ref{fig5} and compared to the
counterpart in natural PbS.}
\label{fig4}
\end{figure}

Data in the isotopically enriched sample have been taken over a smaller
energy range comprising only 3 CCD windows with longer integration times ($%
15\,\mathrm{min}\times 3$), in order to obtain the frequency of the peaks
with better accuracy. The measurement thus requires $45\,\mathrm{min}$ of
integration time for each temperature. The temperature dependence of the
high-wavenumber feature in Pb$^{34}$S can be seen in Fig. \ref{fig4}.

\begin{figure}[h]
 \centering{
  \includegraphics[width=6cm,angle=0]{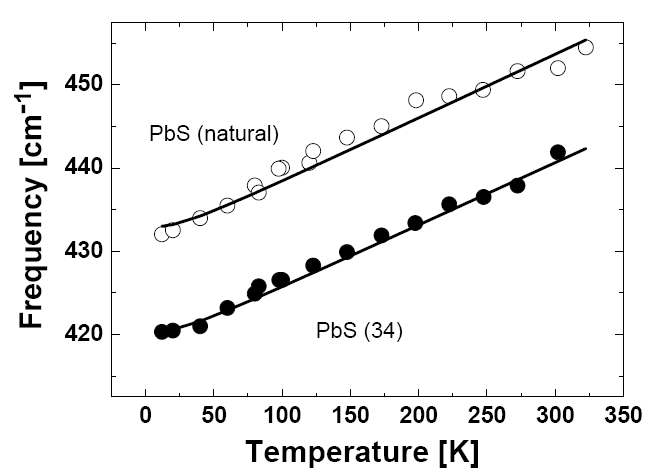}
 }
\caption{Temperature dependence of the main peak in the high-wavenumbers feature of the second-order scattering of both natural and
isotopically enriched Pb$^{34}$S. The solid lines are fits with Eq.
\ref{Eq1}. See the text for further details.}
 \label{fig5}
\end{figure}

Of the two components of this doublet, the one with the highest intensity is
the easiest and most reliable to follow as a function of temperature. The
second peak, at higher frequencies, becomes very weak above $\sim 40\,%
\mathrm{K}$ and its energy cannot be very accurately determined. We thus
concentrate on the temperature dependence of the former which is shown in Fig. \ref{fig5} for both the natural and the isotopically modified samples.

The data shown in Fig. 5 for both samples exhibit the characteristic
temperature dependence expected from anharmonic interactions (the phonon
under consideration decaying into two virtual ones). We have used for the
fits described below both the data displayed here and those reported in Ref. \onlinecite{Sherwin} corresponding to higher temperatures (which overlap only in the $\sim 77-120\,{\rm K}$ temperature range
with our data)\cite{comment}. The Raman shift $\omega$ as a
function of temperature $(T)$ has been fitted with:

\begin{equation}
\omega =\omega _{0}+\Delta \omega \left[ \frac{2}{e^{T_{0}/T}-1}+1\right] ,
\label{Eq1}
\end{equation}%
where $T_{0}$ is a characteristic (average) energy (in units of temperature)
of each of the two virtual phonons responsible for the main (or average)
anharmonic interaction with each of the two ``bare'' frequencies adding
up to the bare $\omega _{0}$ (i.e. the wavenumber that would be observed if
anharmonic effects were not present), $\Delta \omega $ is the \textit{%
frequency renormalization} at $T=0$, i.e. the zero-point anharmonic
renormalization (the real part of the corresponding self-energy). For $T\rightarrow 0$, $\omega \rightarrow (\omega _{0}+\Delta \omega )$,
whilst if $T\gg T_{0}$, $\omega \propto T$ which is the classical limit. If $4T_{0} < \omega _{0}$ ($\omega _{0}$ and $T_{0}$ in the same units) $\Delta \omega$ is positive (because of the sign of the energy denominator in the expression
for the real part of the self-energy\cite{review}). Figure \ref{fig5} shows
the best fits to the data with Eq. \ref{Eq1} for both natural PbS and Pb$%
^{34}$S. The fit parameters are summarized in Table \ref{table}. These
parameters have, actually, large error bars because the wavenumbers of the
peaks change by $\sim 10\,\mathrm{cm}^{-1}$ over the full temperature range
and are hard to pin down to better than $\sim 1\,\mathrm{cm}^{-1}$, as
can be easily surmised by visual inspection of the raw data. Based in the information discussed below (\textit{vide infra}) regarding the dispersion relations and the phonon density of states, the following
picture of the anharmonic interaction responsible for the temperature shift
of the $\sim 430\,\mathrm{cm}^{-1}$mode emerges. The shift is affected by
virtual decay into two modes (65 cm$^{-1})$ with basically Pb and very
little S character. It is therefore reasonable to fix $T_{0}$ to be the same
for both samples, as was done when performing the fit. This decreases the
number of parameters of the combined fits by one and produces more
physically meaningful values for the other parameters. Our fits are
compatible with a change of $\sim 3\%$ and $\sim 2.4\%$ in $\omega _{0}$ and 
$\Delta \omega $, respectively, with the sign\ of these shifts corresponding to $%
\omega _{0}$ and $\Delta \omega $ decreasing with increasing mass, as
expected (see next section).

\begin{table}[tbp]
\centering           
\begin{tabular}{|c|c|c|}
\hline
Parameter & PbS(natural) & Pb$^{34}$S \\ \hline
$\omega_0$ (cm$^{-1}$) & 428.0 & 415.5 \\ \hline
$\Delta\omega$ (cm$^{-1}$) & 5.01 & 4.88 \\ \hline
$T_0$ (in cm$^{-1}$) & 65 & 65 \\ \hline
\end{tabular}%
\caption{Parameters of Eq. \ref{Eq1} as fitted to the data in Fig. \ref{fig5}.
Typical error bars for these parameters are estimated to be $\sim 0.5\,{\rm cm}^{-1}$. See the text for further details.}
\label{table}
\end{table}

\section{Theory}

The interpretation of the phenomenology just described is helped by
calculations of the phonon dispersion relations of PbS and the relevant
densities of one and two-phonon states (the latter with $\mathbf{k}$%
-vectors equal in magnitude but opposite in sign). Semiempirical
calculations are available in the literature\cite{Upadhyaya,Elcombe}; they were performed by fitting a few parameters to a number of experimental frequencies obtained vs. \textbf{k} by inelastic neutron
scattering (INS), as reported in Ref. \onlinecite{Elcombe}. Here we use parameter-free
dispersion relations derived {\it ab initio} using the ABINIT code\cite{ABINIT}. Details of the calculations are given in Ref. \onlinecite{Romero}, which also contains the density of one-phonon states and its
projection on the two constituent atoms of PbS. The corresponding densities
of two-phonon states with \textbf{k}$_{1}=\pm$\textbf{k}$_{2}$ (sum as well
as difference of the two phonons), were not presented in Ref. \onlinecite{Romero}, and
are given here. The ``sum'' processes are responsible for the
anharmonic down-conversion into two phonons of lower energy\cite{Madelung};
while the ``difference'' anharmonic decay describes the process by which a
phonon is destroyed simultaneously with a second low frequency phonon (which
must be already in the crystal) and a phonon of higher energy (with respect
to the initial one) is created. The difference processes tend to have less
importance at low temperatures and vanish for $T\rightarrow 0$. Sum and
difference processes are not only responsible for anharmonic phonon decay
but should also determine the second-order Raman spectra discussed here.

\begin{figure}[tbp]
\centering{\ \includegraphics[width=8cm, height = 6cm,angle=0]{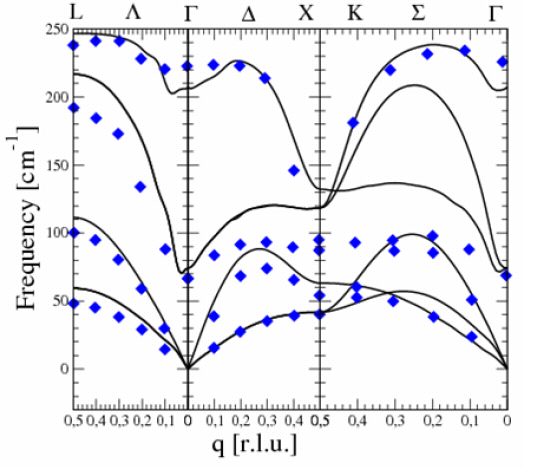} }
\caption{\textit{Ab initio} phonon band structure calculation of PbS along
the principal directions in the Brillouin zone (in reciprocal lattice units: r.l.u). The points represent
experimental determinations from neutron scattering\cite{Elcombe}. Note that some of the
bands have a much flatter dispersion than that predicted in the model. We
shall use experimental points whenever possible and only assign energies to
specific bands when the agreement between theory and experiment is good.}
\label{fig6}
\end{figure}

Figure \ref{fig6} shows the \textit{ab initio} calculated dispersion
relations of PbS along three principal directions of the Brillouin zone
([100], [111] and [110]). The experimental points obtained with INS are
represented by diamonds. The discrepancies which exist between experimental
and calculated points are not surprising when one considers that no
adjustable parameters have been used in the latter. The largest
discrepancies are found for the TO bands. They may be due to anharmonic
shifts which affect the experimental data but are not included in the purely
harmonic calculations. They are in part included in the semiempirical
calculation because of the use of adjustable parameters. One can
nevertheless surmise that the \textit{ab initio} calculations should
provide an overall better agreement to the \textit{harmonic} dispersion
relations than the semiempirical ones. Figures \ref{fig7} and \ref{fig8}
display the sum and difference DOS obtained through Brillouin zone
integration from the dispersion relations of Fig. \ref{fig6}. \ The results in
Figs. \ref{fig6}-\ref{fig8} set the framework for the following discussion.

\section{Comparison with the experiment and discussion}

The dispersion relations in Fig. \ref{fig6} show large regions where the $TA$
bands are rather flat. They give rise to a high and rather narrow peak at $\sim$60\,cm$^{-1}$ (see Fig. 2 in Ref. \onlinecite{Romero}). This peak is likely
to correspond to the frequency $T_{0}$=65\,cm$^{-1}$ given in the Table. In
Fig. 2 of Ref. \onlinecite{Romero}, the partial contribution of the lead and sulfur vibrations (i.e., the eigenvector
components) to the one-phonon DOS is also given. The
large mass difference between these two atoms produces a nearly complete
frequency separation of the corresponding vibrations: up to $\sim 120$\,cm$^{-1}$ the Pb atomic displacements dominate whereas the vibrations above 120\,cm$^{-1}$ are basically dominated by the motion of S atoms. Hence the two phonons contributing to the 
$\sim 430$\,cm$^{-1}$ band should have a squared vibrational amplitude
proportional to (M$_{S})^{-1/2}$ (where M$_{S}$ is the atomic mass of the
corresponding sulfur isotope). The change of (M$_{S})^{-1/2}$ between the two
isotopes considered here (natural S is nearly pure $\ ^{32}$S) is 3\%, a
value which agrees with the fitted variation in \ $\Delta \omega $ ($\Delta
\omega $=2.6\%).

\bigskip

\begin{figure}[h]
\centering{\ \includegraphics[width=8cm, height = 6cm]{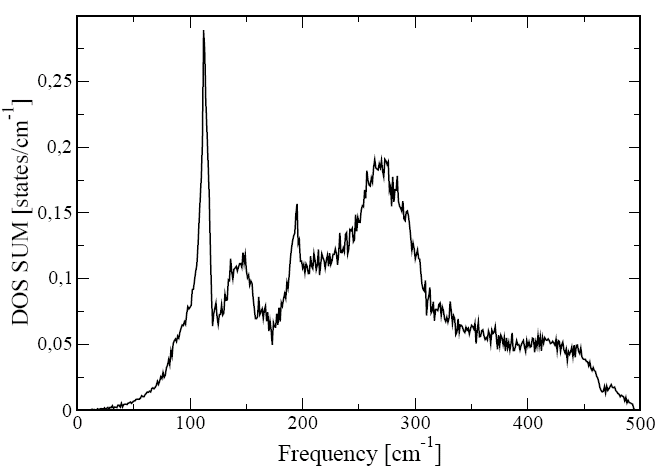} }
\caption{Phonon sum density of states (DOS) calculated from the dispersion relations in
Fig. \ref{fig6}. High wavenumber features related to the LO optical branches have predominantly ``S'' character, while low wavenumber ``acoustic-like'' features have mainly ``Pb'' character. See the text for
further details.}
\label{fig7}
\end{figure}

\begin{figure}[h]
\centering{\ \includegraphics[width=8cm, height = 6cm]{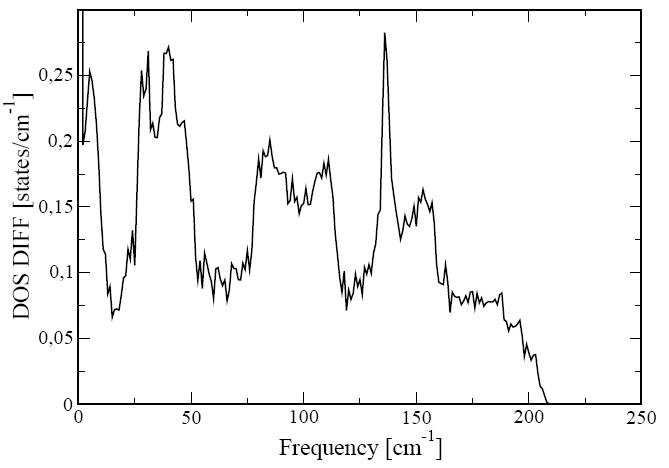} }
\caption{Two phonon difference density of states (DOS) calculated from the
bands in Fig. \ref{fig6}. The difference spectra should occur at the limit of
what we can measure at low frequencies in our experimental setup and should
also have a distinct temperature dependence disappearing for $T\rightarrow 0$
(which is not observed experimentally on the low energy limit of our
observation window). The experimental evidence available at this stage
suggests that the observed spectra are dominated by two-phonon sum processes.}
\label{fig8}
\end{figure}

The densities of states for the sum and difference of two phonons, as
obtained from the $ab$ $initio$ dispersion relations, are displayed in Figs. \ref{fig7} and \ref{fig8}. Multiplied by the appropriate Bose-Einstein factors and coupling constants they determine both anharmonic
self energies and two phonon Raman scattering intensities. The sharp
two-phonon peak seen in Fig. \ref{fig7} at $\sim$120\,cm$^{-1}$ corresponds to the two
phonons (each at T$_{0}$ =60\,cm$^{-1}$) involved in the virtual anharmonic
decay of the 430\,cm$^{-1}$ Raman excitations. Several factors need to be
considered when comparing second-order Raman spectra with phonon sum or
difference DOS. In general, the second-order spectra will resemble (in part)
the sum DOS, i.e. only contributions of two phonons , with $+${\bf k} and $-${\bf k} wave vectors, respectively, and either the same (overtones) of
different (combinations) frequencies. This is, in fact, how the calculated
sum DOS was generated: it contains both overtones and combinations. Although there are many more combinations (5$\times $6=30) than overtones (6) for each {\bf k}, the overtones seen are often enhanced by the
coupling constants (i.e. matrix elements)\cite{LSSII}. A detailed
comparison of the Raman spectra with the two-phonon DOS is not
justified without the introduction of possibly resonant optical matrix
elements and anharmonic self-energies. However, peaks and other structures
observed in the two-phonon DOS may find their counterparts in features
observed in the Raman spectra. The sets of phonons responsible for the
calculated features can be identified by comparison with the dispersion
relations. Anharmonic shifts can then be introduced by taking into account
the difference between calculated and INS-measured phonon frequencies shown
in Fig. \ref{fig6}.

Besides the main $\sim$430\,{\rm cm}$^{-1}$ band, there is also in the measured
spectra the already mentioned $\sim$210\,{\rm cm}$^{-1}$ band, (note that its frequency decreases
with increasing $T$'s\cite{Sherwin}, i.e. the usual behavior for the shifts
of phonons with temperature) and a few other weak bands already reported in Refs. \cite{Sherwin,7,Cao}.

We make the following tentative assignments of both the main and secondary
features:

\begin{itemize}
\item {The experimental curves have a small (but discernible) feature at
about $\sim 160\,\mathrm{cm}^{-1}$. This is at the low energy limit of
our experimental range and therefore can only be assigned tentatively. It
could correspond to $TA+TO$ (combination band) along $\Sigma $ ($\sim
(55+100)\,\mathrm{cm}^{-1}=155\,\mathrm{cm}^{-1}$). The frequency of this
spectral feature also decreases with increasing temperatures\cite{Sherwin}.}

\item {The experimental peak in the region $205-210\,\mathrm{cm}^{-1}$ seems to have a shoulder at $\sim 230\,\mathrm{cm}^{-1}$ seen clearly at low
temperatures. This feature was previously attributed to $LO$ scattering
(which is formally forbidden, but can be made allowed and observable by Fr\"{o}hlich interaction when the laser frequency is near an interband resonance\cite{7}). The data in Refs. \cite{Sherwin,7}, which were not taken at low
temperatures, did not allow a clear identification of the shoulder at about $\sim 220\,\mathrm{cm}^{-1}$ which could be due to two-phonon scattering. In fact,
both features at $\sim 205-210\,\mathrm{cm}^{-1}$ could correspond to
two-phonon scattering, and it is hard to distinguish this assignment from
that to a forbidden $LO$ scattering situation. The sharp peak at $\sim 200\,\mathrm{cm}^{-1}$ in the calculated sum DOS corresponds to $2TO$ phonons
either along $\Delta $ or $\Sigma$. One could also take along $\Sigma $ the 
\textit{experimental} $TO+LA$ values and obtain $\sim 190\,\mathrm{cm}^{-1}$%
. In addition, $2TO$ at both $\Delta $ and $\Sigma $ should give a feature
at $\sim 200\,\mathrm{cm}^{-1}$. With suitable weighting by matrix elements,
all these assignments may play a role in the Raman feature observed
around 210 cm}$^{-1}$.

\item {Combination scattering of $LA+LO$ along $\Sigma$ would give (from the
experimental values) $\sim (100+230)\,\mathrm{cm}^{-1}=330\,\mathrm{cm}^{-1}$%
, or $\sim 340\,\mathrm{cm}^{-1}$ along $\Sigma $. This phonon combination could be responsible for the observed experimental weak structure at $\sim 350\,\mathrm{cm}^{-1}$.}

\item {The broad peak in the calculated sum DOS centered at $\sim 270\,%
\mathrm{cm}^{-1}$ could be twice the (theoretical) low $TO$ along sigma and
have no connection with the experimental phonons (there are no INS points
around $\sim 135\,\mathrm{cm}^{-1}$ and therefore we do not expect a feature
at $270\,{\rm cm}^{-1}$ in the experiment). The actual $TO$ seems to be
lower, possibly because of the anharmonic shift.}

\item {Finally, the asymmetric broad experimental band at $\sim 430\,\mathrm{%
cm}^{-1}$, whose temperature dependence we discussed in the
previous section (Fig. \ref{fig5}), could correspond to (using experimental
values (INS) for the phonon energies of the bands): $(i)$ $2LO$ at $\Delta
~=2\times 230\,\mathrm{cm}^{-1}=460\,\mathrm{cm}^{-1}$; $(ii)$ $2LO$ at $%
\Lambda ~=2\times 245\,\mathrm{cm}^{-1}=490\,\mathrm{cm}^{-1}$; and/or $%
(iii) $ $LO+TO$ at $L$ and $\Lambda ~=(190+240)\,\mathrm{cm}^{-1}=430\,%
\mathrm{cm}^{-1}$}

\item {For the two-phonon difference DOS in Fig. \ref{fig8} there is less
experimental evidence. Unlike the sum DOS of Fig. \ref{fig7}, the intensity
of the features in the difference spectra should vanish at low
temperature. Any peak whose intensity vanishes as $T$ approaches the lowest
possible temperature (12\,K in our case) is a strong candidate to
be compared with the difference DOS. We do not see any feature that could
safely be assigned to the difference DOS, which exhibits strong structures at
30, 90 and 140\,cm$^{-1}$. All these structures lie below the lower limit
of the spectral range that can be recorded with our CCD detector without
being ''swamped'' by stray light from the laser. The sharp peak in Fig. 4 of Ref. \cite{Cao} could be assigned to the peak seen at the same frequency in Fig. \ref{fig8} (due to $LA-TA$ phonons at $X$ and $\Delta$) if it is
confirmed that it belongs to PbS. The best way to buttress this conjecture
would be to measure its intensity as the temperature is lowered. Unfortunately, Ref. \cite{Cao} only reported data obtained at room temperature and our spectrometer does not operate below $\sim 135\,{\rm cm}^{-1}$.}

\item {Another aspect we have already touched upon concerns the
interpretation of the unusual frequency up-shift of the $\sim 430\,\mathrm{cm%
}^{-1}$ feature in Fig. \ref{fig5} due to virtual anharmonic decay into two
phonons. The sign of this coefficient, corresponding to a ''blue'' shift, is
related to the fact that the two-phonon intermediate states are at lower
frequency than each of  the phonons under  consideration ($2\times 60\,{\rm cm}^{-1}=120\,{\rm cm}^{-1}$ vs. $\sim 215\,{\rm cm}^{-1}$. The $\sim 215\,{\rm cm}^{-1}$ modes are all S-like, whereas the 60\,cm$^{-1}$ phonons are lead-like. It has been mentioned in the previous section that this
leads to a zero-point anharmonic renormalization $\Delta \omega $ proportional to \textit{M}$_{S}^{-1/2}$. The corresponding renormalizations for the two isotopes measured are given in the Table. They
differ by 2.6\%, a value which agrees with that determined from the
corresponding isotopic masses of sulfur. For the $205\,\mathrm{cm}^{-1}$
peak, regardless of whether it is due only to forbidden one-$LO$ scattering
or to two phonon scattering, each with a frequency around 100\,cm$^{-1}$, most of the spectral weight of the two phonons with which it interacts lies above the frequency of the phonons being considered (see Fig. \ref{fig7}). There is some experimental evidence for this down-shift in our data, (see
also Ref. \onlinecite{7}) even though the resolution is poor and the exact nature of
the $\sim 210\,\mathrm{cm}^{-1}$ peak in terms of one-phonon forbidden
scattering or two phonon scattering cannot be unambiguously decided at this
stage. Our analysis, however, paves the way for the interpretation of future
experimental observations.}

\item{The two components of the $\sim 430\,{\rm cm}^{-1}$ band must be sulfur
like, according to Figs. \ref{fig6} and \ref{fig7}. Hence they should shift (according to
the M$^{-1/2}$ law) by 3\% from natural PbS to Pb$^{34}$S. The measured
shift, as listed in the Table, is also 3\%. We note that we have estimated a
somewhat smaller shift (2\%) for the measured spectra of the $\sim 210\, \mathrm{cm}^{-1}$ band. Although the difference between this shift and that of the square roots of the sulfur masses (3\%) lies probably within the
experimental error, it would be easy to justify it in terms of the theoretical expectation: For the
one-$LO$ scattering the shift should be determined by the reduced mass of Pb and
S and amount to 2.6\%. In the case of scattering by two phonons, with
frequencies around $\sim 100\, \mathrm{cm}^{-1}$, which fall in the region which
separates Pb-like from S-like vibrations, a shift smaller than that
predicted from the sulfur mass can be expected but it is difficult to
estimate quantitatively.}

\end{itemize}

\section{CONCLUSIONS }
We have measured the Raman spectra of two PbS samples, one made out of natural galena and the other a synthetic crystal with natural sulfur replaced by $^{34}$S, in the 150-500\,cm$^{-1}$ spectral region at temperatures between 10 and 120\,K. These data extend earlier data on the same materials covering the range 100-300\,K. For the quantitative interpretation of the spectra we have performed {\it ab initio} calculations of phonon dispersion relations and the corresponding densities of two-phonon states. The main spectral feature, a broad doublet centered at $\sim 430\,{\rm cm}^{-1}$, has been assigned to scattering by two $LO$ phonons. Its ``anomalous'' temperature dependence has been attributed to virtual anharmonic decay into two phonons of frequency  $\sim 65\,{\rm cm}^{-1}$ each. The dependence of the frequencies of the two main observed features on the isotopic mass of sulphur has been shown to agree with theoretical predictions, a fact which we believe buttresses our assignments and interpretation.

\section{Acknowledgements}
P.G.E acknowledges partial support from the Royal Society of New Zealand
(RSNZ) through a Marsden Grant. A.H.R. has been supported by CONACyT M\'{e}xico under the project J-42647-F, and by a grant from the University of California Institute for M\'{e}xico and the United States (UC MEXUS) and the Consejo Nacional de Ciencia y Tecnolog\'{\i}a de M\'{e}xico (CoNaCyT).


\begin{thebibliography}{99}
\bibitem{comment} There is a $\sim 5\,\mathrm{cm}^{-1}$ rigid shifts in
absolute frequencies between the ones determined here and those in Ref. \onlinecite{Sherwin}; none of our conclusions depends on this difference
(which presumably comes from a calibration difference between the two
different spectrometers located in the northern hemisphere and down under)
and we have simply shifted the data points from Ref. \onlinecite{Sherwin}
to get the best possible overlap with the present data .
\end{thebibliography}
\end{document}